\documentclass[preprint,amsmath,amssymb,prd,superscriptaddress]{revtex4}
\usepackage[dvips]{graphicx}
\usepackage{color}
\usepackage{multirow}
\usepackage{array}
\usepackage{amsmath,amssymb,amsfonts,amstext,amsthm}
\usepackage{algorithm,algorithmic}
\newcommand{\Dodwf}{\mathcal{D}}
\newcommand{\p}{\partial}
\newcommand{\pslash}{p\kern-1ex /}
\newcommand{\lslash}{l\kern-1ex /}
\newcommand{\kslash}{k\kern-1ex /}
\newcommand{\dslash}{\p\kern-1.2ex /}
\newcommand{\Dslash}{{\cal D}\kern-1.5ex /}

\newcommand{\tr}{{\rm tr}}

\newcommand{\CH}{{\cal H}}
\newcommand{\bea}{\begin{eqnarray}}
\newcommand{\eea}{\end{eqnarray}}

\newcommand{\nn}{\nonumber\\}

\newcommand{\BAN}{\begin{eqnarray*}}
\newcommand{\EAN}{\end{eqnarray*}}
\begin{document}

\newcommand{\NTU}{
  Department of Physics,
  National Taiwan University, Taipei~10617, Taiwan
}

\newcommand{\CQSE}{
  Center for Quantum Science and Engineering,
  National Taiwan University, Taipei~10617, Taiwan
}

\newcommand{\CTS}{
  Center for Theoretical Sciences,
  National Taiwan University, Taipei~10617, Taiwan
}

\preprint{NTUTH-14-505A}

\title{Exact Pseudofermion Action for Monte Carlo Simulation of Domain-Wall Fermion}

\author{Yu-Chih~Chen}
\affiliation{\NTU}

\author{Ting-Wai~Chiu}
\affiliation{\NTU}
\affiliation{\CQSE}

\collaboration{TWQCD Collaboration}
\noaffiliation

\pacs{11.15.Ha,11.30.Rd,12.38.Gc}

\begin{abstract}

We present an exact pseudofermion action for hybrid Monte Carlo simulation (HMC) of one-flavor domain-wall fermion (DWF),  
with the effective 4-dimensional Dirac operator equal to the optimal rational approximation 
of the overlap-Dirac operator with kernel $ H = c H_w (1 + d \gamma_5 H_w)^{-1} $, where $ c $ and $ d $ are constants. 
Using this exact pseudofermion action, we perform HMC of one-flavor QCD, 
and compare its characteristics with the widely used rational hybrid Monte Carlo algorithm (RHMC). 
Moreover, to demonstrate the practicality of the exact one-flavor algorithm (EOFA), we perform 
the first dynamical simulation of the (1+1)-flavors QCD with DWF.

\end{abstract}

\maketitle

Quantum Chromodynamics (QCD) is the fundamental theory
for the interaction between quarks and gluons.
It provides the theoretical framework to understand 
the nuclear force/energy from the first principles.
Moreover, QCD plays an important role in the
evolution of the early universe, from the quark-gluon
phase to the hadron phase. Since quarks are relativistic fermions,
they possess the chiral symmetry in the massless limit.
At zero temperature, the chiral symmetry [$ SU_L(N_f) \times SU_R(N_f) $]
of $ N_f $ massless quarks is spontaneously broken to
$ SU_V(N_f) $, due to the strong interaction between
quarks and gluons in the vacuum. This gives the (nearly) massless Goldstone
bosons (pions) and their specific interactions.
To investigate the spontaneous chiral symmetry breaking
as well as hadron physics from the first principles of QCD,
it requires nonperturbative methods.
So far, lattice QCD is the most promising approach,
discretizing the continuum space-time on a 4-dimensional lattice \cite{Wilson:1974sk},
and computing physical observables by Monte Carlo simulation \cite{Creutz:1980zw}.
However, in lattice QCD, formulating lattice fermion
with exact chiral symmetry at finite lattice spacing is rather nontrivial. This
is realized through domain-wall fermions (DWF) on the 5-dimensional lattice \cite{Kaplan:1992bt}
and the overlap-Dirac fermion on the 4-dimensional lattice \cite{Neuberger:1997fp,Narayanan:1994gw}.

Consider the overlap-Dirac operator with bare quark mass $ m_q $, 
\bea
\label{eq:D_mq} 
D = m_q + \frac{(1- r m_q)}{2r} [1+ \gamma_5 H (H^2)^{-1/2} ], \hspace{2mm}  r = 1/[2m_0(1-dm_0)], \hspace{2mm} m_0 \in (0,2).
\eea
Its eigenmodes consist of complex conjugate pairs, and (for topologically non-trivial gauge field) real
eigenmodes with definite chiralities at $ m_q $ and $ 1/r $ satisfying the chirality sum rule \cite{Chiu:1998bh}, 
$ n_+ - n_- + N_+ - N_- = 0 $, 
where $ n_{\pm} $ ($ N_{\pm} $) denote the number of eigenmodes at $ m_q $ ($ 1/r $) with $ \pm $ chirality.
Empirically, the real eigenmodes always satisfy either ($ n_- = N_+ = 0, n_+ = N_- $) or ($ n_+ = N_- = 0, n_- = N_+ $).
Thus, one can write 
\BAN
\det D = \left\{ \begin{array}{ll} 
(r m_q)^{n_+} \det \CH_-^2 = (r m_q)^{-n_+} \det \CH_+^2, & \hspace{2mm} n_+ \geq 0,  \\ 
(r m_q)^{n_-} \det \CH_+^2 = (r m_q)^{-n_-} \det \CH_-^2, & \hspace{2mm} n_- \geq 0,   
                 \end{array}\right.
\EAN
where $ \CH_\pm^2 = P_\pm (D^\dagger D) $, and $ P_\pm = (1 \pm \gamma_5 )/2 $. 
It follows that the pseudofermion action for one-flavor 
overlap fermion can be expressed in terms of $ n_\pm $ and $ \CH_\pm^2 $ (Hermitian and positive-definite), thus is amenable to   
HMC \cite{Duane:1987de}, as studied in Refs. \cite{Bode:1999dd,Fodor:2003bh,DeGrand:2006ws}.
However, this approach requires the computation of the change of $ n_\pm $ at each step of the molecular dynamics in HMC, 
which is prohibitively expensive for large lattices (e.g., $ 16^3 \times 32 $).  
Moreover, the discontinuity of the fermion determinant at the topological boundary highly suppresses  
the crossing rate between different topological sectors, thus renders HMC failing to sample all topological sectors ergodically.
These difficulties can be circumvented as follows.
Firstly, as shown in Ref. \cite{Ogawa:2009ex},  
any positive Dirac operator satisfying $ \gamma_5 $-hermiticity ($ \gamma_5 D \gamma_5 = D^\dagger $) possesses 
a positive-definite pseudofermion action for one-flavor fermion, without explicit dependence on $ n_\pm $. 
Secondly, the step function of the fermion determinant at the topological boundary can be smoothed out by 
using DWF with finite $ N_s $ (e.g., $ N_s = 16 $), 
then the HMC on the 5-dimensional lattice can sample all topological sectors ergodically 
and also keep the chiral symmetry at a good precision (e.g., the residual mass less than 5\% of the bare quark mass). 
This has been demonstrated for 2-flavors QCD in Ref. \cite{Chiu:2011dz}, and for (1+1)-flavors QCD in this paper.

The construction of positive-definite pseudofermion action for HMC of one-flavor DWF has been given in Ref. \cite{Ogawa:2009ex},  
for the conventional DWF with the effective 4-dimensional Dirac operator equal to the polar approximation 
of the overlap-Dirac operator with kernel $ H=\gamma_5 D_w ( 2 + D_w)^{-1} $, 
and for the optimal domain-wall fermion (ODWF) \cite{Chiu:2002ir} with the effective 4-dimensional Dirac operator 
equal to the optimal rational approximation of the overlap-Dirac operator with kernel $ H_w = \gamma_5 D_w $.  
In this paper, we generalize the construction to ODWF 
with the overlap kernel $ H = c H_w (1 + d \gamma_5 H_w)^{-1} $, where $ c $ and $ d $ are constants.
We note that this kernel is the most general form one can have for ODWF, as shown in Ref. \cite{Chen:2012jya}.
Using the exact pseudofermion action, we perform HMC of one-flavor QCD, 
and compare its characteristics with those of RHMC \cite{Clark:2006fx},  
the most widely used algorithm for handling one-flavor fermion in lattice QCD.
Moreover, to demonstrate the practicality of the exact one-flavor algorithm (EOFA), we perform 
the first dynamical simulation of the (1+1)-flavors QCD with ODWF.  

  

In general, the 5-dimensional lattice Dirac operator of all variants of DWF 
\cite{Shamir:1993zy, Borici:1999zw, Chiu:2002ir, Brower:2004xi} can be written as \cite{Chen:2012jya}
\bea
\label{eq:D_odwf}
[\Dodwf(m)]_{xx';ss'} &=& 
  (\rho_s D_w + 1)_{xx'} \delta_{ss'}
 +(\sigma_s D_w - 1)_{xx'} L_{ss'},       
\eea
where $ x $ and $ x' $ denote the lattice sites on the 4-dimensional lattice, $ s $ and $ s' $
the indices in the fifth dimension, and the Dirac and color indices have been suppressed.
Here $D_w$ is the standard Wilson Dirac operator plus a negative parameter $-m_0 \; (0 < m_0 < 2)$,
\BAN
\label{eq:Dw}
(D_w)_{xx'} &=& \sum_{\mu=1}^4 \gamma_\mu (t_\mu)_{xx'} + W_{xx'} - m_0 \delta_{x,x'}, \nn
(t_\mu)_{xx'} &=& \frac{1}{2} \left[ U_\mu(x)\delta_{x+\hat{\mu},x'} - U^\dagger_\mu(x')\delta_{x-\hat{\mu},x'} \right], \\
W_{xx'} &=& 4 \delta_{x,x'} - \frac{1}{2} \sum_{\mu=1}^4 \left[ 
                                         U_\mu(x)\delta_{x+\hat{\mu},x'}+U^\dagger_\mu(x')\delta_{x-\hat{\mu},x'} \right], 
\EAN
%
where $U_\mu(x)$ denotes the link variable pointing from $ x $ to $ x + \hat\mu $.  
The operator $ L $ is independent of the gauge field, and it can be written as  
\BAN
L = P_+ L_+ + P_- L_-, \quad P_\pm = (1\pm \gamma_5)/2,
\EAN
and
\BAN
\label{eq:L}
(L_+)_{ss'} = (L_-)_{s's}= \left\{ 
    \begin{array}{ll} 
      - m \delta_{N_s,s'}, & s = 1, \\  
      \delta_{s-1,s'}, & 1 < s \leq N_s,   
    \end{array}\right.
\EAN
where $ N_s $ is the number of sites in the fifth dimension,    
$ m \equiv r m_q $, $m_q $ is the bare quark mass, and $ r = 1/[2m_0(1-dm_0)] $.
Note that the matrices $ L_{\pm} $ satisfy $ L_\pm^T = L_\mp $, and $ R_5 L_\pm R_5 = L_\mp $, 
where $ R_5 $ is the reflection operator in the fifth dimension, with elements $ (R_5)_{ss'} = \delta_{s',N_s+1-s} $.
Thus $ R_5 L_\pm $ is real and symmetric.

Different ways of assigning the values of $ \rho_s $ and $ \sigma_s $     
along the fifth dimension give all variants of DWF. 
In general, we write $ \rho_s = c \omega_s + d $, and $ \sigma_s = c \omega_s - d $, where $ c $ and $ d $ are constants.
For the conventional DWF with the Shamir kernel \cite{Shamir:1993zy}, $c=d=1/2 $, and $ \omega_s = 1, \forall s $.
For the Borici DWF \cite{Borici:1999zw}, $ c=1 $, $ d=0 $, and $ \omega_s = 1, \forall s $.
For the M\"obius DWF \cite{Brower:2004xi}, $ \omega_s = 1, \forall s $.
For the optimal DWF, the weights $ \{ \omega_s \} $ are fixed according to the formula derived in \cite{Chiu:2002ir},   
then its effective 4-dimensional Dirac operator is exactly equal to the Zolotarev optimal rational approximation \cite{Chiu:2002eh}
of the overlap-Dirac operator (\ref{eq:D_mq}). 

Since the matrices $ L $ and $\omega = \mbox{diag}(\omega_1, \cdots, \omega_{N_s})$ are independent of the gauge field,
we can drop the factor $ [ d + c \omega (1+L)(1-L)^{-1} ] $ from the DWF operator (\ref{eq:D_odwf}) and obtain the 
re-scaled DWF operator for HMC, 
\bea
\label{eq:DT}
D_{T}(m) 
\equiv D_w+P_+M_+(m) +P_- M_-(m),
\eea
where  
\bea
\label{eq:M_pm}
M_{\pm}(m) &=& \omega^{-1/2}[\omega^{-1}d + c N_{\pm}(m)]^{-1}\omega^{-1/2},\\
\label{eq:N_pm}
N_{\pm}(m) &=& [1+L_{\pm}(m)][1-L_{\pm}(m)]^{-1}. 
\eea
Here the dependence on $ m \equiv r m_q $ has been shown explicitly in $ L_{\pm} $, $ M_\pm $, and $ N_\pm $. 
Using the relation 
\BAN
N_\pm(m) = N_\pm(0) - \frac{2m}{1+m} u u^T, \quad u^{T} \equiv (1,1,\cdots,1), 
\EAN
and the Sherman-Morrison formula, we obtain   
\bea
\label{eq:Mpm}
[\omega^{-1}d+cN_{\pm}(m)]^{-1}   
= \left[A_{\pm}-\frac{2cm}{1+m}uu^{T}\right]^{-1} 
= A_{\pm}^{-1} + \frac{2cm}{1+ m - 2 c m \lambda_\pm } A_{\pm}^{-1}uu^{T} A_{\pm}^{-1}, 
\eea
where  
\BAN
A_{\pm} &\equiv& \omega^{-1}d+cN_{\pm}(0), \\
\lambda_\pm &\equiv& u^T A_\pm^{-1} u.
\EAN
Now we use $\omega$ which is invariant under $ R_5 $,  
i.e., $ R_5 \omega R_5 = \omega $, define $v_{\pm} \equiv R_5 A_{\pm}^{-1} u$, and put (\ref{eq:Mpm}) into (\ref{eq:M_pm}), 
then we obtain 
\bea
\label{eq:Mpm_vvT}
M_{\pm}(m) = \omega^{-1/2}A_{\pm}^{-1}\omega^{-1/2} + 
             \frac{2cm}{1+m-2cm \lambda_\pm } R_5 \omega^{-1/2} v_{\pm}v_{\pm}^{T}\omega^{-1/2}.
\eea
We note that the reflection-symmetric $ \omega $ is different from 
the $ \omega $ given in Ref. \cite{Chiu:2002ir}, and the details will be given in a forthcoming paper.

Since $ A_{\pm}^{-1} $ is an upper/lower triangular matrix, we can solve $ v_{\pm} $ exactly 
with the following recursion relation,  
\bea
(v_+)_{N_s} &=& (v_-)_{1} = \alpha_{N_s},\\
(v_+)_s &=& (v_-)_{N_s-s+1} = \alpha_s \beta_{s+1} (v_+)_{s+1}, \quad s=N_s-1, \cdots, 1, 
\eea
where $\alpha_s \equiv 1/(\omega_s^{-1}d+c)$ and $\beta_s \equiv \omega_s^{-1}d-c$.
Then we obtain
\bea
\lambda_{-} = \lambda_{+} = u^{T}A_{+}^{-1}u = u^{T} R_5 A_{+}^{-1}u = u^{T}v_{+} = \sum_s (v_+)_s = \sum_s \alpha_s Q_s \equiv \lambda, 
\eea
where $ Q_s \equiv \alpha_{s+1}\beta_{s+1}...\alpha_{Ns}\beta_{Ns} $.

In the following, without loss of generality, we use the Dirac matrices in the chiral representation,  
\BAN
\gamma_{\mu} =
\left(
\begin{array}{cc}
       0            &  \sigma_{\mu} \\
\sigma_{\mu}^{\dagger} &        0
\end{array}
\right), \hspace{4mm}
\sigma_{\mu} = \left( \vec{\sigma} , i I \right), \hspace{4mm}
\gamma_{5} = \gamma_1 \gamma_2 \gamma_3 \gamma_4 = 
\left(
\begin{array}{cc}
       I    &   0  \\
       0    &  -I
\end{array}
\right), \hspace{4mm}
\EAN
where $\vec{\sigma}$ are the Pauli matrices. 
Next, we define 
\bea
\label{eq:DTm1m2}
D_T(m_1,m_2) \equiv 
\left(
\begin{array}{cc}
      W-m_0+M_+(m_1)      &  \sigma \cdot t \\
-(\sigma\cdot t)^{\dag} &   W-m_0+M_-(m_2)
\end{array}
\right), 
\eea
which is equal to $ D_T(m) $ [Eq. (\ref{eq:DT})] when $m_1=m_2=m$. 
After incorporating the contribution of the Pauli-Villars fields, the fermion determinant of the DWF becomes 
$ \det D_T(m) / \det D_T(1) $. Using the Schur decompositions, we obtain
\bea
\label{eq:DT1overDTm}
\frac{\det D_T(1)}{\det D_T(m)}
  = \frac{\det[W-m_0+M_-(1)]^2 \cdot \det H_+(1)}{\det[W-m_0+M_+(m)]^2 \cdot \det H_-(m)},
\eea
where   
\bea
\label{eq:Hp}
H_+(m_1,m_2) &\equiv& R_5\left[ W-m_0+M_+(m_1) + (\sigma \cdot t) \frac{1}{W-m_0+M_-(m_2)}(\sigma \cdot t)^\dagger \right],  \\
\label{eq:Hm}
H_-(m_1,m_2) &\equiv& R_5\left[ W-m_0+M_-(m_2) + (\sigma \cdot t)^\dagger \frac{1}{W-m_0+M_+(m_1)}(\sigma \cdot t) \right],  
\eea
which become $ H_+(m) $ and $ H_-(m) $ when $ m_1 = m_2 = m $.
Since $ R_5 \omega R_5 = \omega $, this implies that 
$ (R_5 M_{\pm})^{\dag} = R_5 M_{\pm} $ and $ (M_{\pm} R_5)^{\dag} = M_{\pm} R_5 $, thus $H_\pm $ is Hermitian. 
Applying the Schur decompositions to $ D_T(m,1) $,  
we obtain 
\bea
\label{eq:DetRelation}
\frac{\det[W-m_0+M_-(1)]^2}{\det[W-m_0+M_+(m)]^2} = \frac{\det[H_-(m)+\Delta_-(m)]}{\det[ H_+(1) - \Delta_+(m)]}, 
\hspace{3mm} \Delta_{\pm}(m) \equiv R_5[M_{\pm}(1)-M_{\pm}(m)].
\eea
Using (\ref{eq:Mpm_vvT}), we obtain
\bea
\Delta_{\pm}(m) = k \omega^{-1/2}v_{\pm}v_{\pm}^{T}\omega^{-1/2} = k \Omega_\pm \Omega^T_\pm, 
\eea
where
\bea
&& k \equiv \frac{c}{1-c\lambda}\frac{1-m}{(1+m-2cm\lambda)}, \\
&& (\Omega_\pm)_{s,s'} \equiv \omega_s^{-1/2} (v_\pm)_s \delta_{s',1}.
\eea 
Substituting (\ref{eq:DetRelation}) into (\ref{eq:DT1overDTm}), we immediately have
\bea
\label{eq:DT1overDTmFinal}
&& \frac{\det D_T(1)}{\det D_T(m)}
= \frac{\det[H_-(m)+\Delta_-(m)]}{\det[H_+(1) - \Delta_+(m)]}\frac{\det H_+(1)}{\det H_-(m)} 
= \det H_1(m) \cdot \det H_2(m), 
\eea
where  
\bea
\label{eq:H1}
H_1(m) &\equiv& I + k \Omega_{-}^T \frac{1}{H_-(m)}  \Omega_{-}, \\
\label{eq:H2}
H_2(m) &\equiv& I + k \Omega_{+}^T \frac{1}{H_+(1) - \Delta_+(m)} \Omega_{+}.
\eea
Here $ H_1 $ and $ H_2 $ are Hermitian operators (with color and 2-spinor indices) on the 4-dimensional space,  
and the formula $ \det (I + AB) = \det(I + BA) $ has been used in the last equality of (\ref{eq:DT1overDTmFinal}). 
It is trivial to assert that $[W-m_0+M_{\pm}(m)]^{-1} $ (in Eqs. (\ref{eq:Hp})-(\ref{eq:Hm})) is well-defined for $ m>0 $, 
and $ H_1 $ and $ H_2 $ are positive-definite, as shown in Ref. \cite{Ogawa:2009ex}.

From (\ref{eq:DT1overDTmFinal}), the pseudofermion action for one-flavor DWF reads
\bea
\label{eq:Spf}
S_{pf} = \phi_1^{\dag} H_1(m) \phi_1 + \phi_2^{\dag} H_2(m) \phi_2,  
\eea
where $ \phi_1 $ and $ \phi_2 $ are pseudofermion fields on the 4-dimensional lattice, each of two spinor components.
However, the operators $ H_1(m) $ and $ H_2(m) $ are not practical since each involves the inverse of some matrix 
which contains the inverse of another matrix. Again, using the Schur decompositions, we finally have  
\bea
\label{eq:Spf_HT}
S_{pf}
&=& \left(0  \hspace{4mm}  \phi_1^{\dag} \right)
        \left[I- k \Omega_{-}^T \frac{1}{H_T(m)} \Omega_{-} \right]
        \left(\begin{array}{c}
           0  \\
        \phi_1
        \end{array}\right)+ \nonumber \\
&  & \left(\phi_2^{\dag}  \hspace{4mm} 0\right)
        \left[I+ k \Omega_{+}^T \frac{1}{H_T(1) - \Delta_+(m) P_+} \Omega_{+} \right]
        \left(\begin{array}{c}
        \phi_2 \\
           0
        \end{array}\right),  
\eea
where $ H_T(m) \equiv \gamma_5 R_5 D_T(m) $ is a Hermitian operator. This is the main result of this paper.

To generate $ \phi_1 $ and $ \phi_2 $ from Gaussian noise fields $ \eta_1 $ and $ \eta_2 $, 
we use Zolotarev optimal rational approximation for the inverse square root of $ H_1(m) $ and $ H_2(m) $, 
\BAN
\phi_1  &=& \frac{1}{\sqrt{H_1}} \eta_1
                = \sum_{l=1}^{N_p} \frac{b_l}{d_l+H_1} \eta_1
                = \sum_{l=1}^{N_p} b_l e_l
                   \frac{1}{I+ e_l k \Omega_{-}^T [H_-(m)]^{-1} \Omega_{-} } \eta_1, \\
\phi_2  &=& \frac{1}{\sqrt{H_2}} \eta_2
                = \sum_{l=1}^{N_p} \frac{b_l}{d_l+H_2} \eta_2
                = \sum_{l=1}^{N_p} b_l e_l
                    \frac{1}{I+ e_l k \Omega_{+}^T [ H_+(1) - \Delta_+(m)]^{-1} \Omega_{+} } \eta_2 ,
\EAN
where $ e_l \equiv 1/(1+d_l) $, and $N_p$ is the number of poles in the Zolotarev approximation.
Further simplifications can be obtained using the Schur decomposition, and the final results are 
\bea
\label{eq:HB_ZA_phi1}
\left(\begin{array}{c}
        \xi_1  \\
        \phi_1
        \end{array}\right)
= \sum_{l=1}^{N_p} \left[  b_l e_l I
                         + b_l e_l^2 k \Omega_{-}^T \frac{1}{H_T(m)- e_l \Delta_-(m)P_-} \Omega_{-} \right]
    \left(\begin{array}{c}
        0  \\
        \eta_1
        \end{array}\right), \nonumber 
        \\
\label{eq:HB_ZA_phi2}
\left(\begin{array}{c}
        \phi_2  \\
        \xi_2
        \end{array}\right)
= \sum_{l=1}^{N_p} \left[  b_l e_l I
                         - b_l e_l^2 k \Omega_{+}^T
                          \frac{1}{H_T(1)- d_l e_l \Delta_+(m)P_+} \Omega_+ \right]
    \left(\begin{array}{c}
        \eta_2  \\
        0
        \end{array}\right), \nonumber 
\eea
where $\xi_1$ and $ \xi_2 $ are irrelevant fields. Thus $ \phi_1 $ and $ \phi_2 $ can be solved by the conjugate gradient.   
Finally we use the accept-reject algorithm to make sure that $ \phi_1 $ and $ \phi_2 $ give 
the pseudofermion action $ S_{pf} $ (\ref{eq:Spf_HT}) such that the probability distribution $ \exp(-S_{pf}) $ 
satisfies exactly the Gaussian distribution $ \exp(-\eta_1^\dagger \eta_1 - \eta_2^\dagger \eta_2) $. 

In the following, we compare EOFA with RHMC. 
For the memory requirement, it is straightforward to obtain 
the following formula for the ratio of the memory consumption of these two algorithms \cite{Chen:2014bb}   
\BAN
\label{eq:memory_ratio}
\frac{M_{\rm RHMC}}{M_{\rm EOFA}} = \frac{20+3(3+ 2 N_p )N_s}{32 + 10.5 N_s},  
\EAN
where $ N_p $ is the number of poles used in the rational approximation of RHMC, and $ N_s $ is the extent in the fifth dimension.
For $ N_p = 12 $ and $ N_s = 16 $, the ratio is 6.58 for any 4D lattices. In other words, if EOFA requires 12 GB 
to perform HMC of lattice QCD with DWF on the $ 32^3 \times 64 \times 16 $ lattice, then RHMC with 12 poles 
needs at least 79 GB to perform the simulation. Obviously, the memory-saving feature of EOFA is crucial 
for large-scale simulations of lattice QCD with GPUs, in view of each GPU 
having enormous floating-point computing power but limited device memory.
For example, using EOFA, two GPUs (each of 6 GB device momory, e.g., Nvidia GTX-TITAN) 
working together with OpenMP/MPI is capable to simulate lattice QCD with $ (u,d,s,c) $ DWF quarks on the 
$ 32^3 \times 64 \times 16 $ lattice (attaining sustained 780 Gflops for two GTX-TITANs).

\begin{figure*}[tb]
\begin{center}
\begin{tabular}{@{}c@{}c@{}}
\includegraphics*[width=8cm,clip=true]{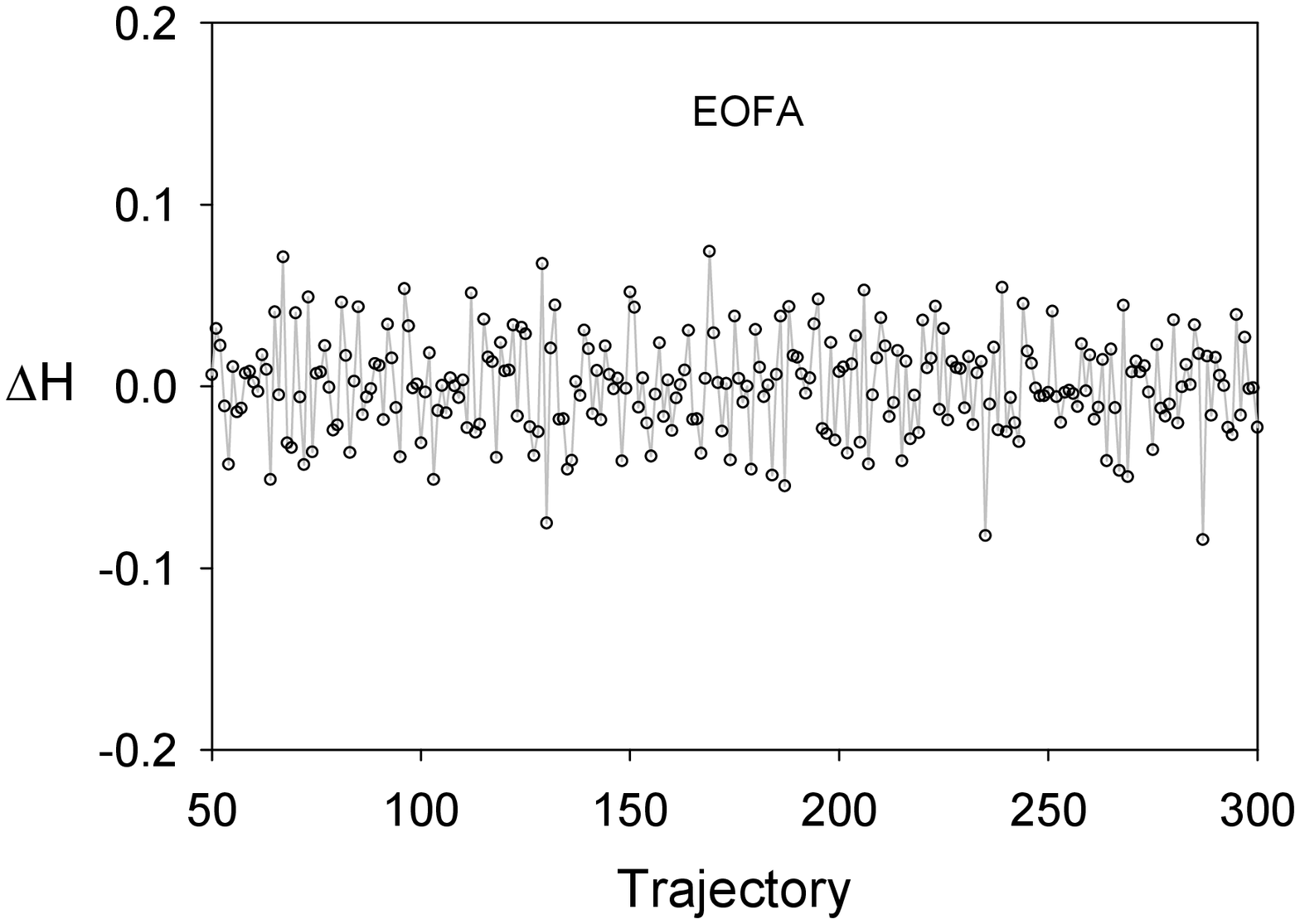}
&
\includegraphics*[width=8cm,clip=true]{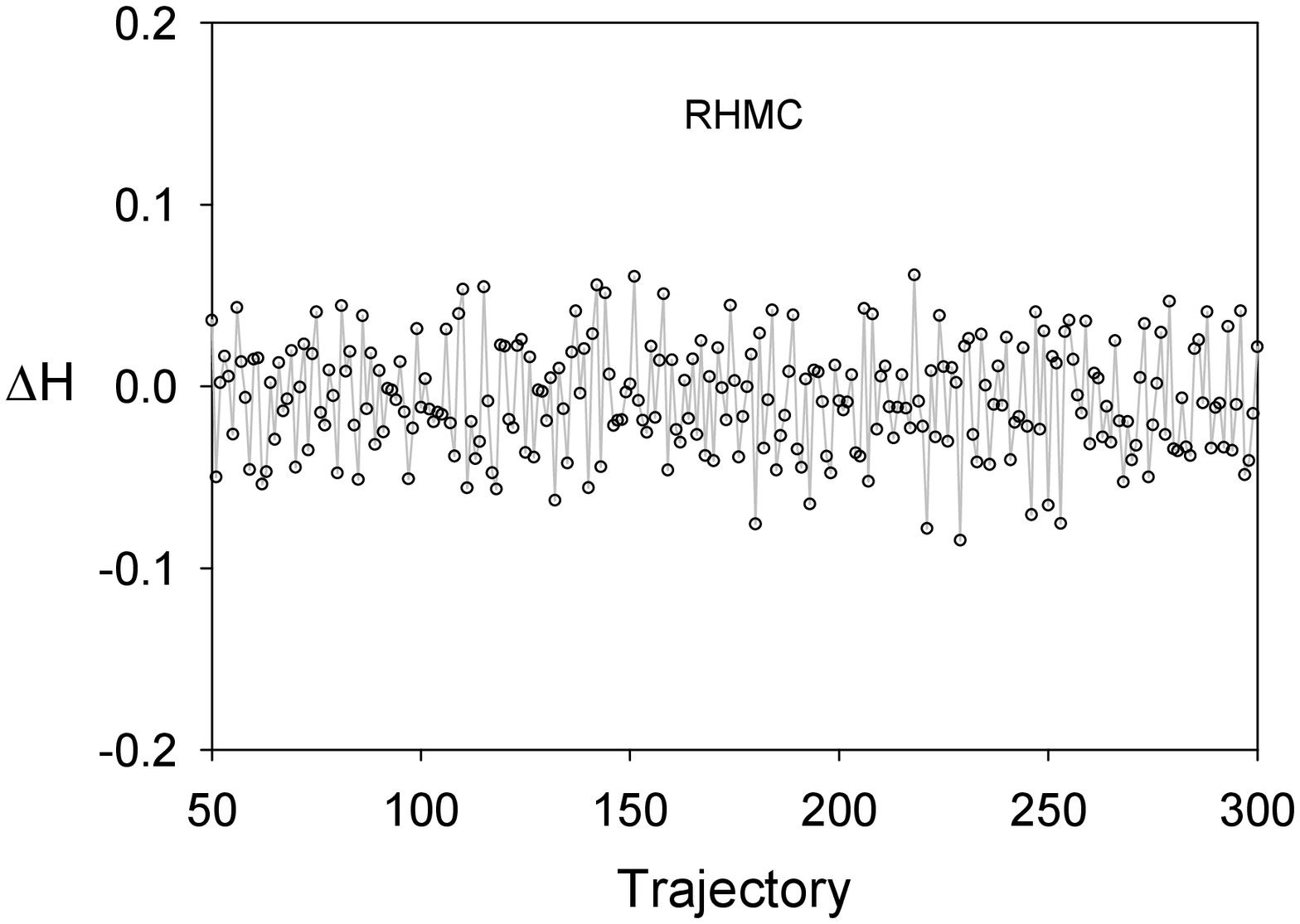}
\\ (a) & (b)
\end{tabular}
\caption{The change of Hamiltonian $ \Delta H $ versus the trajectory in the HMC of one-flavor QCD 
         with the conventional DWF, for (a) EOFA, and (b) RHMC respectively.
         The line connecting the data points is only for guiding the eyes.}
\label{fig:DeltaH}
\end{center}
\end{figure*}

\begin{figure*}[tb]
\begin{center}
\begin{tabular}{@{}c@{}c@{}}
\includegraphics*[width=8cm,clip=true]{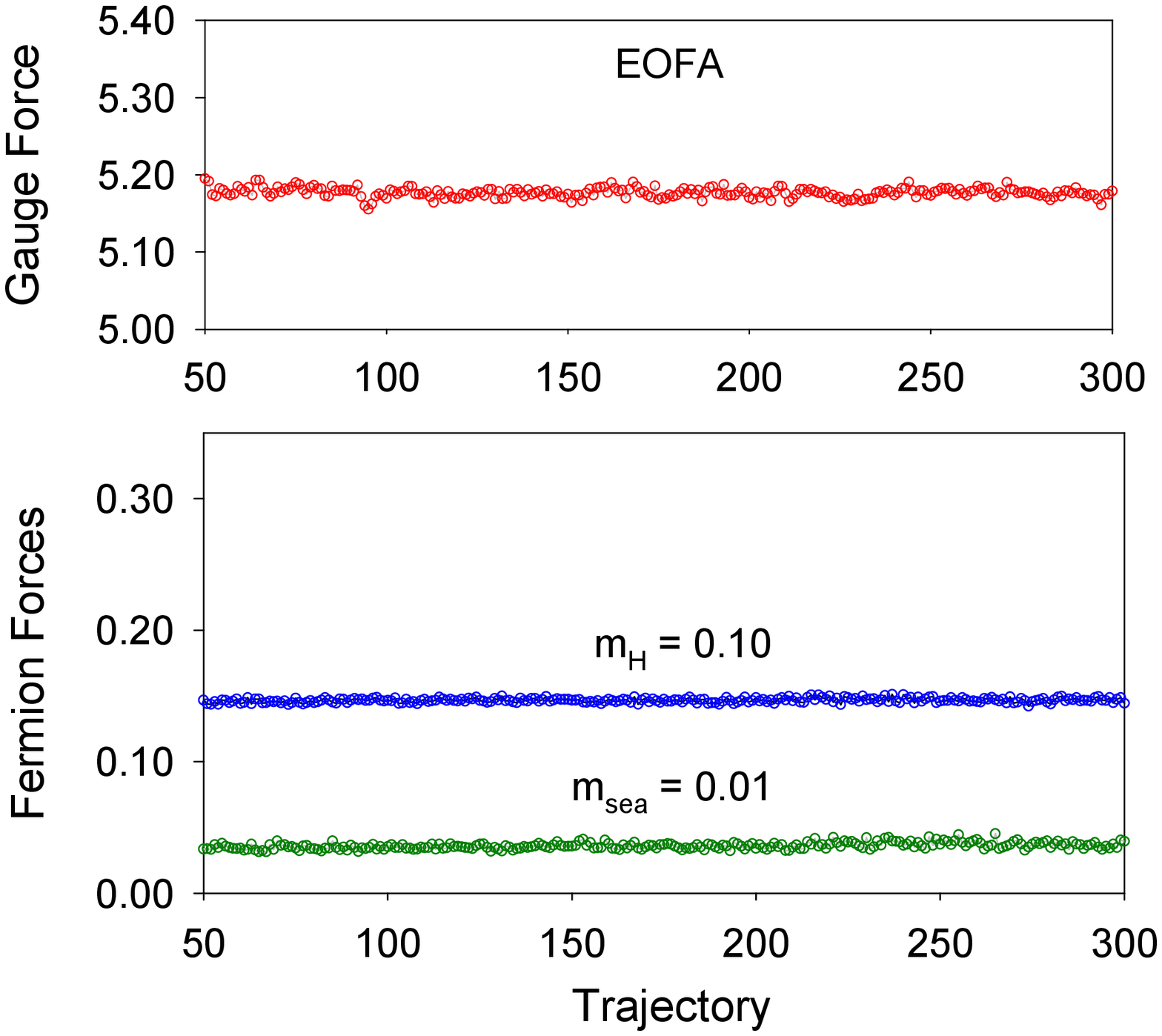}
&
\includegraphics*[width=8cm,clip=true]{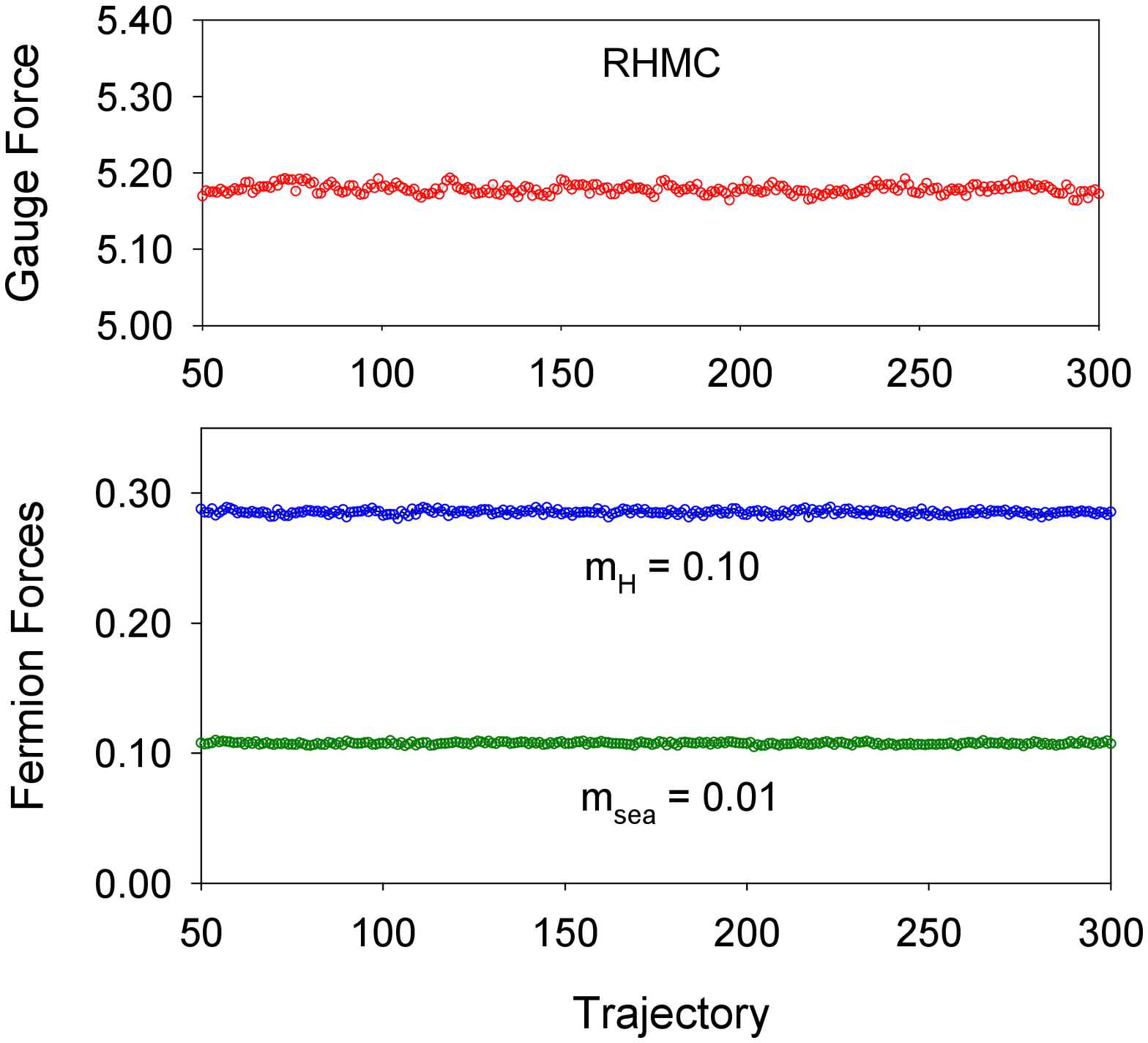}
\\ (a) & (b)
\end{tabular}
\caption{The maximum forces of the gauge field, heavy fermion field, and light fermion field  
         versus the trajectory in the HMC of one-flavor QCD with the conventional DWF, 
         for (a) EOFA, and (b) RHMC respectively.}
\label{fig:HMC_forces}
\end{center}
\end{figure*}

To compare the HMC characteristics of EOFA and RHMC,  
we perform HMC of one-flavor QCD on the $ 8^3 \times 16 $ lattice, 
with the conventional DWF at $ N_s = 16 $ and $ m_0 = 1.8 $, sea-quark mass $ m_{sea} a = 0.01 $, 
and the Wilson plaquette gauge action at $ \beta = 5.95 $. 
In the molecular dynamics, we use the Omelyan integrator \cite{Omelyan:2001}, 
auxillary heavy fermion field \cite{Hasenbusch:2001ne} with $ m_H a = 0.1 $, and multiple-time scale method \cite{Sexton:1992nu}.
The pseudofermion action for Monte Carlo simulation of one-flavor QCD with RHMC is  
\BAN
S_{pf}^{N_f=1} = \phi^\dagger (C_1^\dagger C_1)^{1/4} (C C^\dagger)^{-1/2} (C_1^\dagger C_1)^{1/4} \phi, 
\EAN
where $ C $ is defined in Eq. (13) of Ref. \cite{Chiu:2013aaa}, and the number of poles used in the optimal rational 
approximation of $ (C C^\dagger)^{-1/2} $ and $ (C_1^\dagger C_1)^{1/4} $ is $ N_p = 12 $.
In Fig. \ref{fig:DeltaH}, we plot the change of Hamiltonian $ \Delta H $
of each trajectory after thermalization, for EOFA and RHMC respectively. 
In both cases, $ \Delta H $ is quite smooth, without spikes in all trajectories.
Moreover, the measured values of $ \left< \exp(-\Delta H) \right> $ are: $ 0.9999(16) $ for EOFA, and  
$ 1.0074(18) $ for RHMC, both in good agreement with the condition $ \left< \exp(-\Delta H) \right> = 1 $
which follows from the area-preserving property of the HMC.
In Fig. \ref{fig:HMC_forces}, we plot the maximum force (averaged over all links)
among all momentum updates in each trajectory, for the gauge field, the heavy fermion field,
and the light fermion field respectively. For both EOFA and RHMC, 
the forces all behave smoothly for all trajectories. However, the fermion forces of EOFA are 
substantially smaller than their counterparts in RHMC.  
Using one core of Intel i7-3820 CPU@3.60GHz, the average time for generating one HMC trajectory 
after thermalization is 6644(43) seconds for EOFA, versus 6629(24) seconds for RHMC. 
Taking into account of the acceptance rate 0.987(7) for EOFA, and 0.997(3) for RHMC,  
both EOFA and RHMC have compatible efficiencies.
Further details of the comparison will be given in Ref. \cite{Chen:2014bb}. 

To demonstrate the practicality of EOFA, we perform the first dynamical simulation of the (1+1)-flavors QCD with DWF,   
which also provides gauge ensembles for studying the isospin symmetry breaking effects in the hadron spectrum as well 
as other physical quantities. In the following, we outline the salient features of our simulation. 
We generate the gauge ensembles on the $16^3 \times 32$ lattice with the Wilson gauge action 
at $ \beta = 6/g^2 = 5.95$ (with lattice spacing $ a \sim 0.1 $~fm), 
for three sets of sea-quark masses: $ (m_u, m_d) = \{ (0.01, 0.02), (0.015, 0.03), (0.02, 0.04) \} $,  
with corresponding charged pion masses in the range 250-330 MeV. 
Here the ratio $ m_d/m_u $ has been fixed to $ 2 $, close to its physical limit. 
For the quark part, we use the optimal domain-wall fermion (ODWF) \cite{Chiu:2002ir}
with $ c = 1, d = 0 $ (i.e., $ H = H_w $), $ N_s = 16 $, and $ \lambda_{min}/\lambda_{max} = 0.05/6.2 $.
For each $ (m_u, m_d) $ sea-quark mass, we generate the initial 300-400 trajectories with a Nvidia GPU.
After discarding the initial 200 trajectories for thermalization, we sample one configuration
every 5 trajectories, resulting 20-32 ``seed" configurations for each $ (m_u, m_d) $ sea-quark mass.
Then we use these seed configurations as the initial configurations for independent simulations on 20-32 GPUs.
Each GPU generates 200-250 trajectories independently.
Then we accumulate a total of $\sim 4500$ trajectories for each $ (m_u, m_d) $ sea-quark mass.
From the saturation of the binning error of the plaquette, as well as
the evolution of the topological charge,
we estimate the autocorrelation time to be around 10 trajectories.
Thus we sample one configuration every 10 trajectories,
and obtain $\sim 450$ configurations for each $ (m_u, m_d) $ sea-quark mass.

\begin{figure}[!htb]
\begin{center}
\begin{tabular}{@{}cccc@{}}
\includegraphics*[height=6.0cm,width=5.5cm,clip=true]{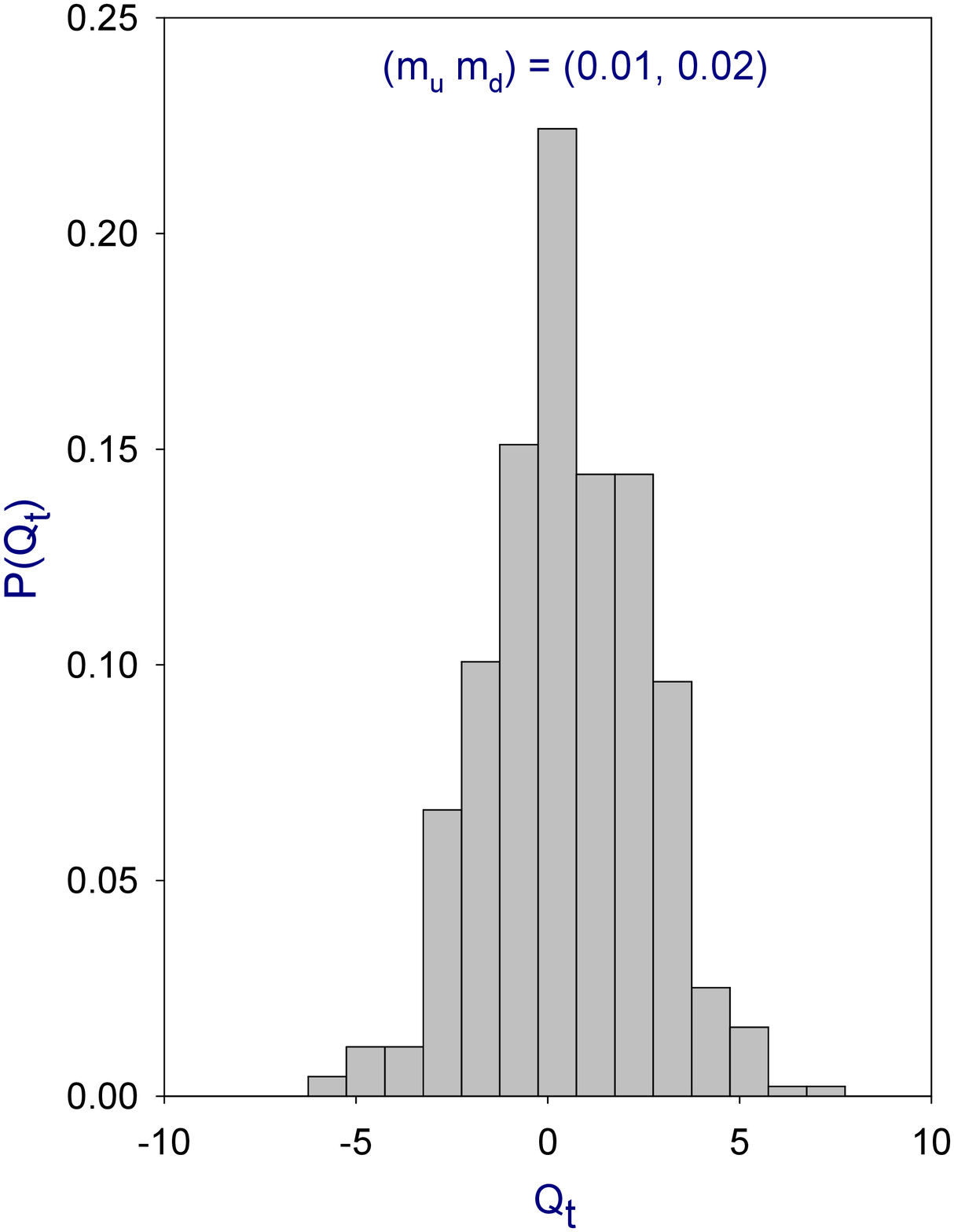}
&
\includegraphics*[height=6.0cm,width=4.5cm,clip=true]{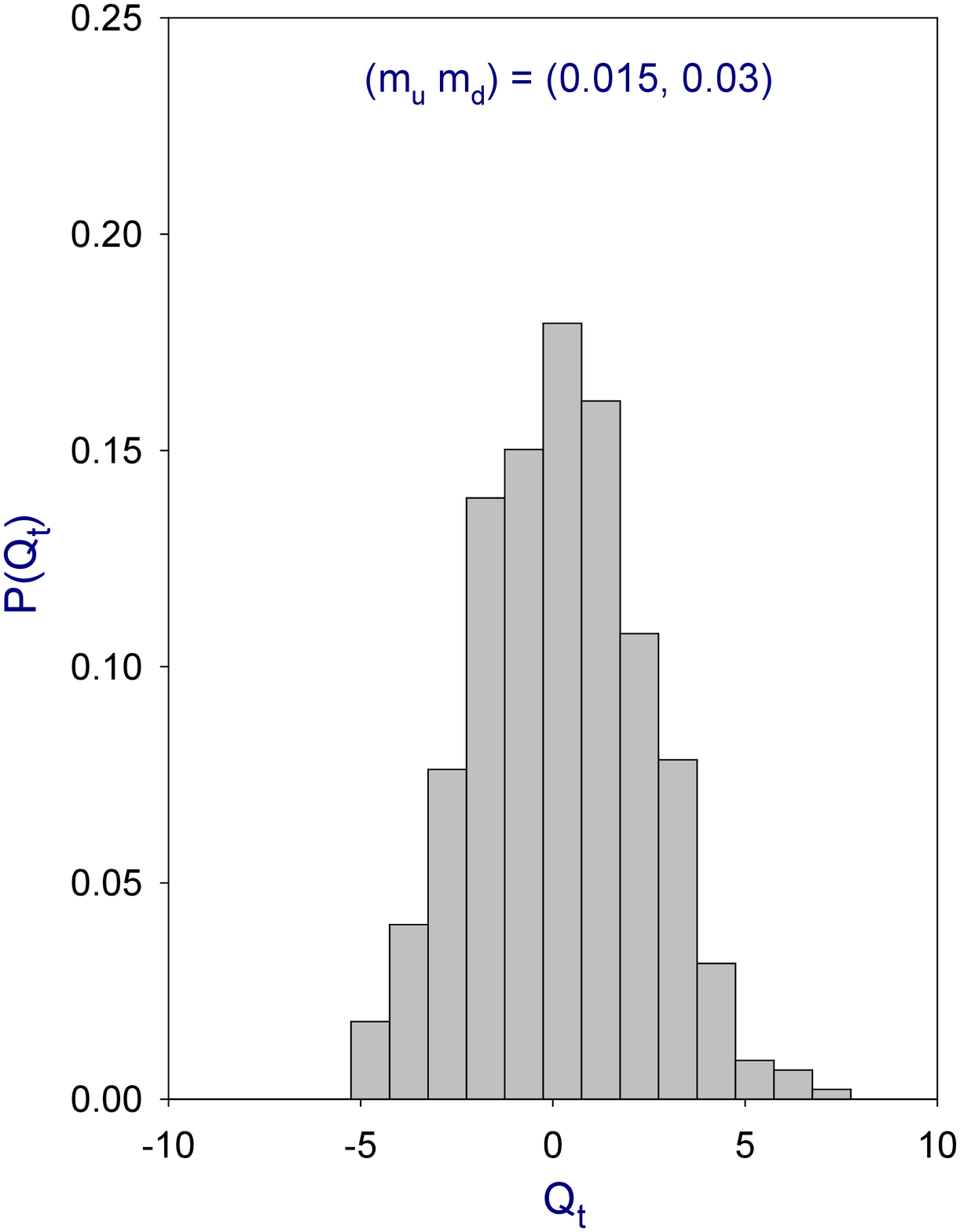}
&
\includegraphics*[height=6.0cm,width=4.5cm,clip=true]{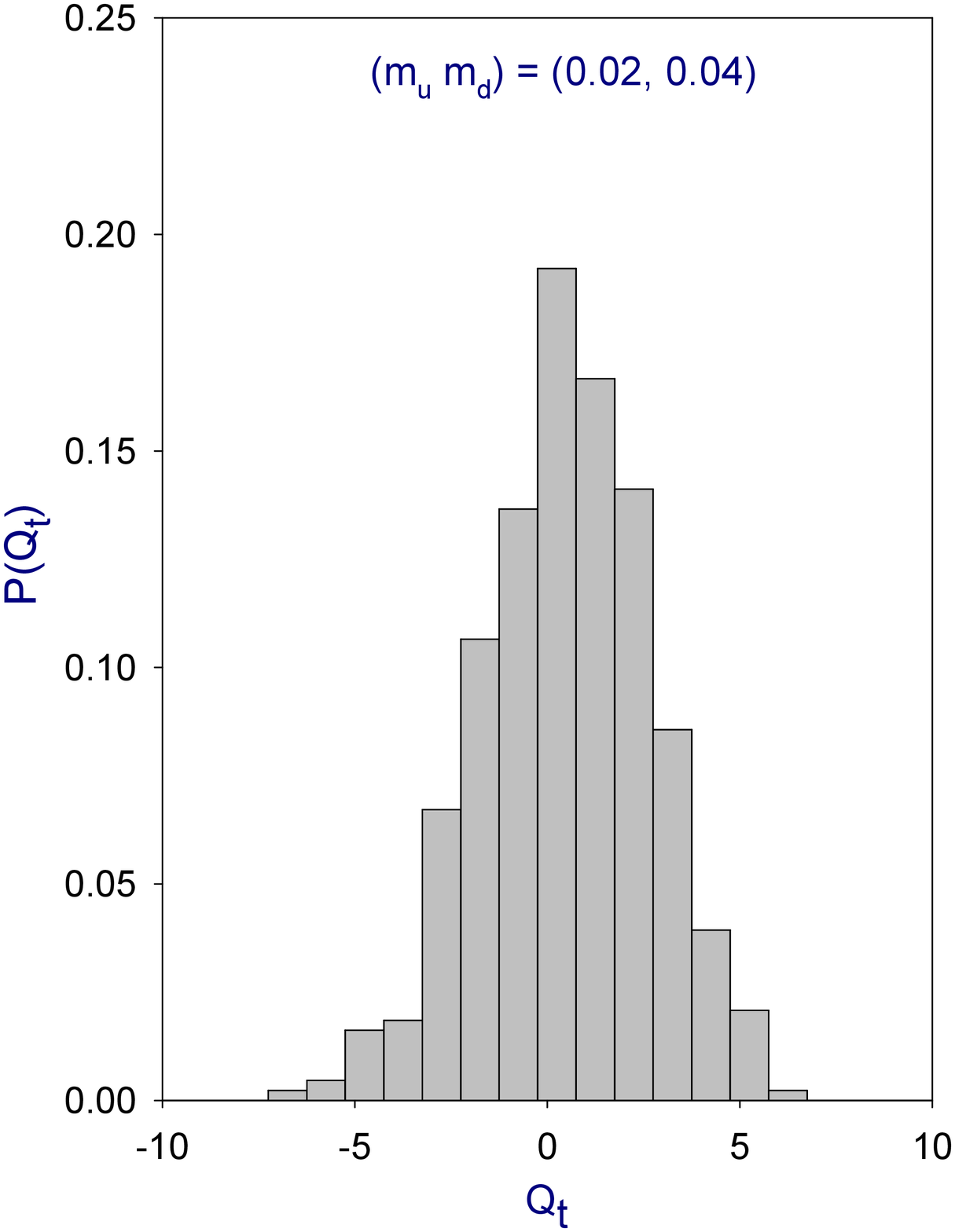}
&
\\
\end{tabular}
\caption{Histogram of topological charge distribution
for three gauge ensembles with $ (m_u, m_d) = \{(0.01,0.02), (0.015, 0.03), (0.02, 0.04) \} $.
}
\label{fig:Q_hist}
\end{center}
\end{figure}

In Fig.~\ref{fig:Q_hist}, we plot the histogram of the topological charge ($ Q_t $)
distribution for these three ensembles.
Evidently, the probability distribution of $ Q_t $ for each ensemble
behaves like a Gaussian, and it becomes more sharply peaked around
$ Q_t = 0 $ as the sea-quark mass gets smaller.
Here the topological charge $ Q_t = \sum_x \epsilon_{\mu\nu\lambda\sigma} \tr[ F_{\mu\nu}(x) F_{\lambda\sigma}(x) ]/(32 \pi^2) $,
where the matrix-valued field tensor $ F_{\mu\nu}(x) $ is obtained from the four plaquettes
surrounding $ x $ on the ($\hat\mu,\hat\nu$) plane. Even though the resulting topological charge
is not exactly equal to an integer, the probability distribution $ P(Q_t) $ suffices to demonstrate that
the HMC indeed samples all topological sectors ergodically.

\begin{figure*}[tb]
\begin{center}
\begin{tabular}{@{}c@{}c@{}}
\includegraphics*[width=7.5cm,clip=true]{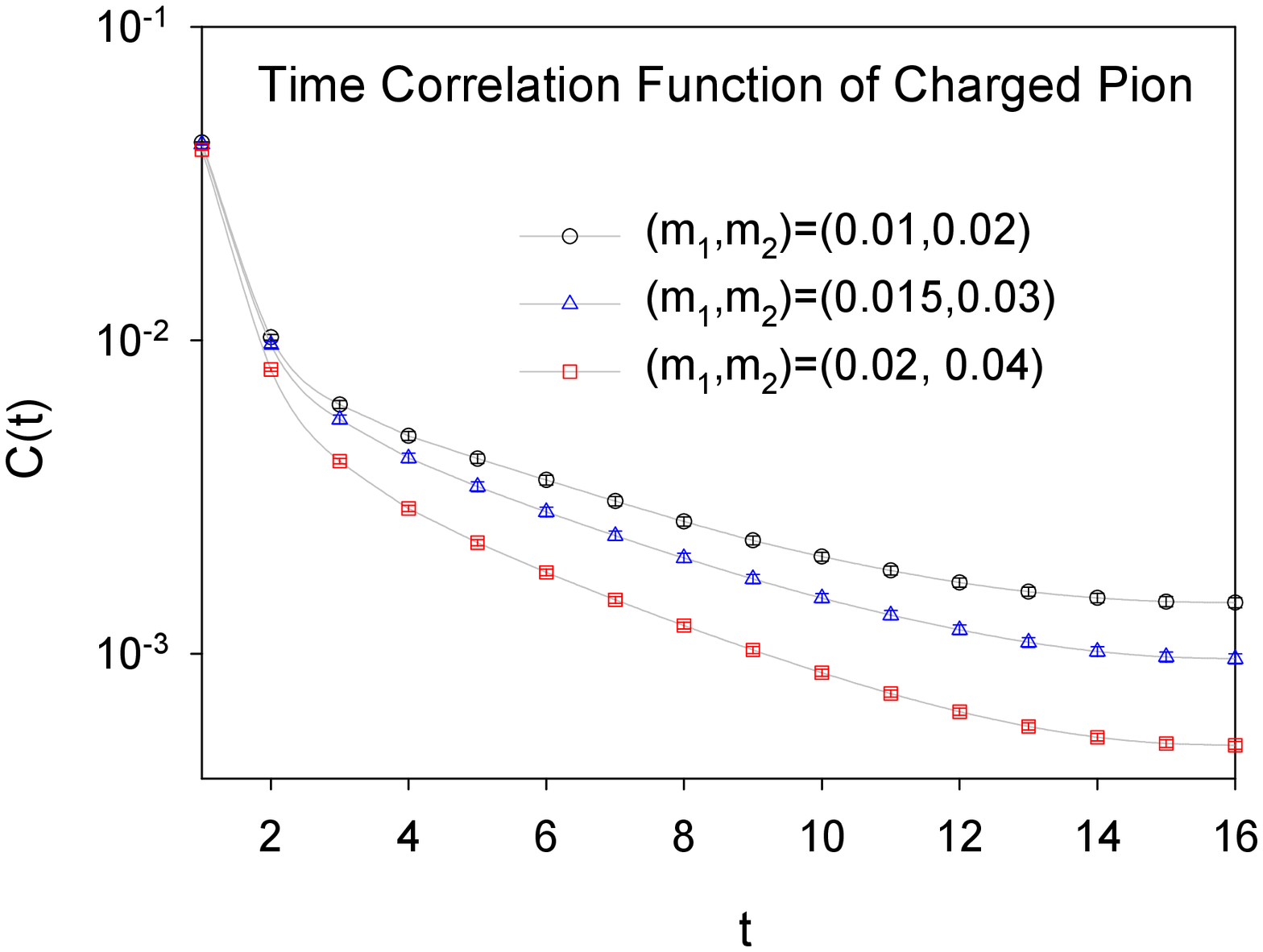}
&
\includegraphics*[width=7.5cm,clip=true]{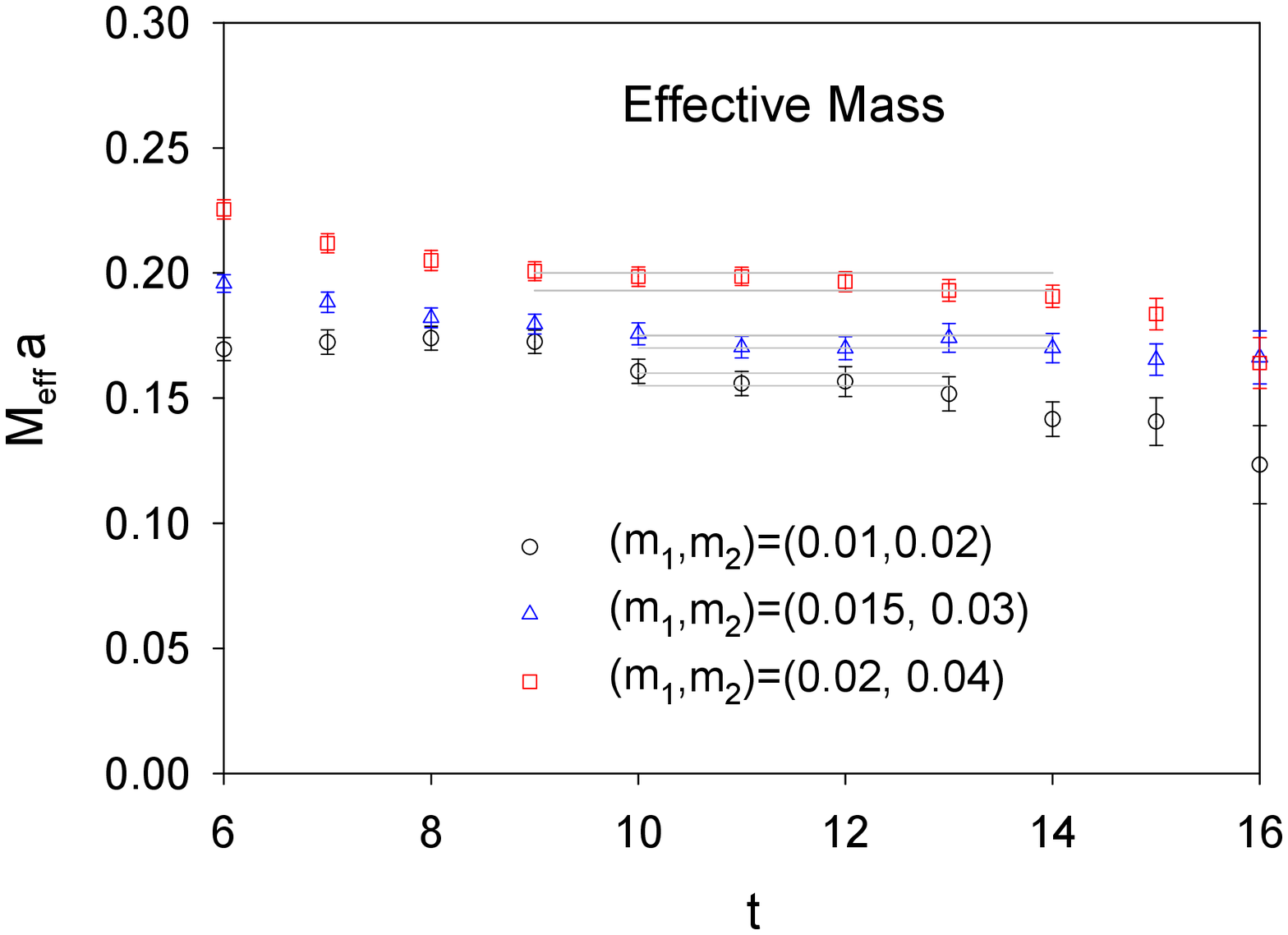}
\\ (a) & (b)
\end{tabular}
\caption{(a) The time-correlation function of charged pion.
         (b) The effective mass of (a).
    Each solid line in (a) connecting the data points of the same $ (m_u, m_d) $ is for guiding the eyes.
    The horizontal lines in (b) denote the fitted masses with the error bars. }
\label{fig:G55a_meff_b595_mu_md}
\end{center}
\end{figure*}

We compute the valence quark propagator with the point source at the origin, and with parameters exactly the same
as those of the sea-quarks ($ N_s = 16 $ and $ \lambda_{min}/\lambda_{max} = 0.05/6.2 $).
For each ensemble, we measure the time-correlation function $ C(t) $ of the charged pion, and fit $ C(t) $  
to the formula $ z^2 [ e^{-M t} + e^{- M (T-t)} ]/(2 M) $ to extract the mass $ M $ and 
the decay constant $ f = (m_u + m_d) z /(2 M^2 ) $. 
In Fig. \ref{fig:G55a_meff_b595_mu_md},
we plot the time-correlation function $ C(t) $ and the effective mass
of the charged pion for $ (m_u, m_d) = \{ (0.01, 0.02), (0.015, 0.03), (0.02, 0.04) \} $.  
Further studies with these three gauge ensembles will be presented in a forthcoming paper.
 
To summarize, we present an exact pseudofermion action for HMC of one-flavor DWF,  
with the effective 4-dimensional Dirac operator equal to the optimal rational approximation 
of the overlap-Dirac operator with kernel $ H = c H_w (1 + d \gamma_5 H_w)^{-1} $, where $ c $ and $ d $ are constants. 
The efficiency of EOFA is compatible with that of RHMC, for the lattices 
($8^3 \times 16 \times 16 $, and $8^3 \times 24 \times 16 $) we have tested so far. 
For larger lattices, we expect that EOFA would outperform RHMC, and the detailed analysis will be given in Ref. \cite{Chen:2014bb}. 
Moreover, the memory consumption of EOFA is much smaller than that of RHMC.  
These features make EOFA a better choice for large-scale simulations of lattice QCD with DWF.   
Finally, we perform the first dynamical simulation of (1+1)-flavors QCD with domain-wall fermion,  
which demonstrates that it is feasible to perform large-scale simulations of lattice QCD with EOFA. 
Now TWQCD Collaboration is using EOFA to simulate lattice QCD with $ (u,d,s,c) $ quarks 
on the $ 24^3 \times 48 \times 16 $ and $ 32^3 \times 64 \times 16 $ lattices, with Nvidia GPUs (GTX-TITAN).

  This work is supported in part by the Ministry of Science and Technology 
  (No.~NSC102-2112-M-002-019-MY3) and NTU-CQSE (Nos.~103R891404).

\end{document}